\documentstyle[12pt]{article}
\begin{document}

\author{$^\star$M. V. Cougo-Pinto\thanks{\it marcus@if.ufrj.br}, $^\star$C. Farina\thanks{\it farina@if.ufrj.br}, $^{\star\P}$J. F. M. Mendes\thanks{\it jayme@if.ufrj.br} e $^\star$A. C. Tort\thanks{\it tort@if.ufrj.br}\vspace{1cm} \\$^\star${\it Instituto de F\'\i sica-UFRJ, CP 68528,}\\{\it Rio de Janeiro, RJ, 21.945-970}\\$^\P${\it 
Instituto de Pesquisa e Desenvolvimento,}\\{\it Rio de Janeiro, RJ, 23.020-470}\\{\it Brazil}}

\date{}

\title{Vacuum confinement at finite temperature for scalar QED in magnetic field and deformed boundary condition}

\maketitle

\begin{abstract}
We investigate the Casimir effect at finite temperature for a charged scalar field in the presence of an external uniform and constant magnetic field, perpendicular to the Casimir plates.  We have used a boundary condition characterized by a deformation parameter $\theta$; for $\theta=0$ we have a periodic condition and for $\theta=\pi$, an antiperiodic one, for 
intermediate values, we have a deformation. The temperature was introduced using the imaginary 
time formalism and both the lagrangian and free energy were obtained from Schwinger proper time 
method for computing the effective action. We also computed the permeability and its asymptotic 
expressions for low and high temperatures. 
\end{abstract}

%%%%%%%%%%%%%%%%%%%%
\vspace{3cm}
%%%%%%%

%%%%%%%%%%%%%
{\small 

The electromagnetic field in classical vacuum is described by Maxwell's lagrangian 
$-F^{\mu\nu}F_{\mu\nu}/4$. When only the field ${\bf B}$ is present, the lagrangian is given by 
${\cal L}^{(0)}({\bf B})=-{\bf B}^2/2$. 

The effective lagrangian, that takes into account the free electromagnetic field and the 
contribution from the medium is described by ${\cal L}_{\rm effective}={\cal L}^{(0)}({\bf 
B})+{\cal L}^{(1)}({\bf B})$, where ${\cal L}^{(1)}({\bf B})$ is the one loop lagrangian of the system. Expanding ${\cal L}^{(1)}({\bf B})$, the term proportional to ${\bf B}^2$ is associated to the linear polarization, while higher order terms represent nonlinear polarizations. 

In 1935, Euler and Heisenberg \cite{heisenberg/1935} obtained the corrections to the effective Lagrangian under an electromagnetic field; the so called Euler-Heisenberg lagrangian. Their result gives the exact expression of the Lagrangian and the first correction is given by a term in $B^4$ \cite{heisenberg/1935}. The correspondent Lagrangian for scalar QED is referred to as the Weisskopf-Schwinger effective lagrangian\cite{weisskopf/1936, schwinger/1951}. 

Consider now a constant and uniform magnetic field directed to ${\cal OZ}$ axis and perpendicular to the Casimir plates. The plates are plane, perfectly conductive, parallel and $a$ is the distance between them. The plates may also be considered squares of side ${\ell}\gg{a}$. 

The boundary condition used here establishes that the scalar field experiences a recoil in phase, at each displacement $a$ along the direction perpendicular to the Casimir plates 
\cite{xxenfpc}. This condition is characterized by $\phi({\bf r}+a{\bf z})=e^{-i\theta}\phi({\bf r})$. For $\theta=0$ or, in the limit $2\pi$, this condition is reduced to the periodic one and for $\theta=\pi$ it is antiperiodic. For intermediate values we 
have a deformed boundary condition. We can consider a circular compactified dimension where the deformed condition occurs, denoted by $S^1_{\theta}$. This condition presents the peculiarity of recoiling the phase of the field in $\theta$ at each complete turn; naturally, for $\theta=0$ we have $S^1_0=S^1$, that is, the common compactification of ${\bf R}$ in $S^1$. In other words, the subjacent space to the scalar field is compactified from ${\bf R}^3$ to ${\bf R}^2\times S^1_{\theta}$,

It can be shown that this boundary condition can be generated from the minimum coupling of the charged scalar field in the spacetime given by ${{\bf R}^2}\times{{\it S}^1}\times{\bf R}$, by the use of a constant potential $\theta/ea$ along the  compactified dimension $S^1$ \cite{xxenfpc}. 

We use here the imaginary time formalism \cite{kapusta} to write down Schwinger's proper-time formalism \cite{schwinger/1951} where the partition function for the bosonic field is 

\begin{equation}
\log{\cal Z}=\frac{1}{2}\int_{s_0}^{\infty}\frac{ds}{s}Tre^{-isH}, 
\end{equation}
where $s$ is the Schwinger's proper-time, $H$ is the proper-time hamiltonian and $s_0$ is a cut off in the proper time $s$.  

The operator $H$ for such charged scalar field is given by $H=(P-eA)^2+m^2$, where $P^2={\bf P}^2-(P^0)^2$. The eigenvalues $p_x$ and $p_y$ are restricted to Landau levels and given by ${p_x}^2+{p_y}^2ë(2n+1)$, where $n\in{\bf N}$. The imposition of the boundary condition on the $z$ component of the momentum gives: $p_z=(2n_1\pi-\theta)/a, n_1\in{\bf Z}$; finally, the Matsubara frequencies are $p^{0}=(2\pi{n_2}i/\beta)$, where $\beta=1/T$ and $n_2\in{\bf Z}$. 

Then the trace reads:

\begin{equation}
Tr\,e^{-isH}=2\frac{eB{\ell}^2}{2\pi}\sum_{n=0}^{\infty}e^{-iseB(2n+1)}\sum_{n_{1}=-\infty}^{\infty}e^{-is(({2n_{1}\pi-\theta})/{a})^2}\sum_{n_{2}=-\infty}^{\infty}e^{is(2n_{2}{\pi}i/\beta)^2}e^{-ism^2} \label{traco}
\end{equation}
where the factor two is due to the degrees of freedom of the complex scalar field and $eB{\ell}^2/2\pi$ comes from the degeneracy of the Landau levels. The sum over the Landau levels is given by $\displaystyle\sum_{n=0}^{\infty}e^{-iseB(2n+1)}={\rm csch}(iseB)/2$ and the sum in $n_1$ and in $n_2$ can be transformed using the Poisson formula. \cite{poisson}: 

\begin{equation}
\sum_{n=-\infty}^{\infty}e^{in^{2}\pi\tau+2inz}=\frac{e^{i\pi/4}}{\sqrt{\tau}}\sum_{n=-\infty}^{\infty}e^{(z-n\pi)^2/i\pi\tau}\label{poisson}.
\end{equation}
The partition function can then be written as:

\begin{eqnarray}
\log{\cal Z}=\frac{{\ell}^2a\beta}{16{\pi}^2}\int_{s_0}^{\infty}\frac{ds}{s^3}(seB){\rm csch}(seB)e^{-sm^2}\left[
1+2\sum_{n_1=1}^{\infty}\cos(n_1\theta)e^{-(n_{1}a)^2/4s}\right]\left[1+2\sum_{n_2=1}^{\infty}e^{-(n_2\beta)^2/4s}\right], 
\end{eqnarray}

where an adequate rotation on complex plane of integration was performed. 

\newpage

Let's now write down the total lagrangian of the system, knowing that ${\cal L}=\log{\cal 
Z}/a{\ell}^2\beta$.

\begin{eqnarray}
{\cal L}=-\frac{B^2}{2}+\frac{1}{16{\pi}^2}\int_{0}^{\infty}\frac{ds}{s^3}e^{-sm^2}(seB){\rm csch}(seB)+ 
\frac{1}{8{\pi}^2}\sum_{n_1=1}^{\infty}\cos(n_1\theta)\int_{0}^{\infty}\frac{ds}{s^3}(seB){\rm csch}(seB)e^{-sm^2-(n_{1}a)^2/4s}+
\nonumber \\
+\frac{1}{8{\pi}^2}\sum_{n_2=1}^{\infty}\int_{0}^{\infty}\frac{ds}{s^3}(seB){\rm csch}(seB)e^{-sm^2-(n_{2}\beta)^2/4s}+
\nonumber\\
+\frac{1}{4{\pi}^2}\sum_{n_1=1}^{\infty}\sum_{n_2=1}^{\infty}\cos(n_1\theta)\int_{0}^{\infty}\frac{ds}{s^3}(seB){\rm csch}(seB)e^{-sm^2-(n_{1}a)^2/4s-(n_{2}{\beta})^2/4s}.\label{L/mu} 
\end{eqnarray}

When we take a close look at this equation we see that the first term gives an infinite contribution: $m^4{\Gamma}(-2)/16\pi^2$ that can be simply subtracted  since it has no physical meaning. The second term of the expansion, proportional to $B^2$, is also divergent but 
relevant because it depends on the external field {\bf B}. Its contribution is given by $-(e^2\Gamma(0)/48\pi^2)(B^2/2)$ that, summed to the Maxwell's lagrangian, becomes the 
renormalisation constant, $Z_3$, \cite{dittrich-reuter}, of the Lagrangian. The complete 
lagrangian is now renormalized and both the field and charge will now be given by 
$B_R=Z_3^{-1/2}B$ and $e_R=Z_3^{1/2}e$. 

Considering a linear magnetization, only terms in $B^2$, the effective lagrangian will be given by:

\begin{equation}
{\cal L}_{\rm effective}=-\frac{1}{2\mu}B^2=-\frac{1}{2}B^2+{\cal L}_{WSC}^{'}(a,B)+{\cal L}_{WS}^{'}(B,\beta)+{\cal L}_{WSC}^{'}(a,B,\beta), 
\end{equation}

and the magnetic permeability reads:

\begin{eqnarray}
\frac{1}{\mu}=1+\frac{e^2}{24\pi^2}\bigg[\sum_{n_1=1}^{\infty}\cos(n_1\theta)\int_{0}^{\infty}\frac{ds}{s}e^{-sm^2-(n_{1}a)^2/4s}+
\sum_{n_2=1}^{\infty}\int_{0}^{\infty}\frac{ds}{s}e^{-sm^2-(n_{2}\beta)^2/4s}+
\nonumber \\
+2\sum_{n_1=1}^{\infty}\sum_{n_2=1}^{\infty}\cos(n_1\theta)\int_{0}^{\infty}\frac{ds}{s}e^{-sm^2-(n_{1}a)^2/4s-(n_{2}\beta)^2/4s}\bigg]. 
\end{eqnarray}

These integrals can be identified as Bessel functions: 
\begin{eqnarray}
\frac{1}{\mu}=1+\frac{e^2}{12\pi^2}\bigg[\sum_{n_1=1}^{\infty}\cos(n_1\theta)K_{0}(amn_1)+
\sum_{n_2=1}^{\infty}K_{0}(\beta{mn_2})
\nonumber \\
+2\sum_{n_1=1}^{\infty}\sum_{n_2=1}^{\infty}\cos(n_1\theta)K_0\left(m\sqrt{(n_1a)^2+(n_2\beta)^2}\right)\bigg].\label{mu-bT} 
\end{eqnarray}

In equation (\ref{mu-bT}), the first term gives the magnetic permeability of the free vacuum at $T=0$. The first sum gives the influence of confinement alone, the second one is the thermal contribution in the case of no confinement and finally, the third sum is a mixed term that gives both confinement and finite temperature contributions. For $a\rightarrow\infty$ the third term vanishes and if $\beta\rightarrow\infty$, this expression takes the form encountered in \cite{guida1} and arbitrary values of $\theta$, $am$ and $\beta{m}$ result in diamagnetic or paramagnetic medium.

For low temperatures, $\beta{m}{\gg}1$, and the permeability will be given by:

\begin{eqnarray}
\frac{1}{\mu}{\approx}1+\frac{e^2}{12\pi^2}\bigg[\sum_{n=1}^{\infty}\cos(n\theta)K_{0}(amn)+
\sqrt{\frac{\pi}{2}}\frac{e^{-{\beta}m}}{\sqrt{\beta{m}}}
+2\sqrt{\frac{\pi}{2}}\cos(\theta)\frac{e^{-m\sqrt{a^2+\beta^2}}}{\sqrt{m\sqrt{a^2+\beta^2}}}\bigg]. 
\end{eqnarray}

To obtain the expression for $1/\mu$ at high temperatures we have to come back to (\ref{traco}) and apply (\ref{poisson}) only for $n_1$. Analogous calculations give for total lagrangian:

\begin{eqnarray}
{\cal L}=\frac{1}{8{\pi}^{3/2}\beta}\int_{0}^{\infty}\frac{ds}{{s}^{5/2}}(seB){\rm csch}(seB)e^{-sm^2}\bigg[1+2\sum_{n_1=1}^{\infty}\cos(n_1\theta)e^{-(n_1{a})^2/4s}+2\sum_{n_2=1}^{\infty}e^{-s(2n_2\pi/\beta)^2}\nonumber \\
+4\sum_{n_1=1}^{\infty}\sum_{n_2=1}^{\infty}\cos(n_1\theta)e^{-(n_1{a})^2/4s-s(2n_2\pi/\beta)^2}\bigg].\label{L-mu2}
\end{eqnarray}

The contribution for $1/\mu$ from the integrals of the first and the third terms between brackets, considering the linear approximation for magnetization are: 
$e^2/24\pi\beta{m}+e^2/24{\pi}^{2}\displaystyle\sum_{n_2=1}^{\infty}1/\sqrt{(m\beta/2\pi)^2+{n_2}^2}$. This sum diverges and must be regularized, what can be done by the Abel-Plana formula \cite{abel-plana}. After doing that, we find $e^2/12\pi^2\displaystyle\sum_{n_2=1}^{\infty}K_0(\beta{mn_2})$, present in (\ref{mu-bT}). We can reach this by choosing an easier way. It is enough to use the (\ref{poisson}) again on the second sum in (\ref{L-mu2}) and later express it in its asymptotic form for $\beta{m}\ll{1}$. The equation for $1/\mu$ is obtained from the effective lagrangian using a procedure similar to what has been already done, and is given by,

\begin{eqnarray}
\frac{1}{\mu}=1+\frac{e^2}{12\pi^2}\bigg[\frac{1}{\beta}\sqrt{\frac{2\pi{a}}{m}}\sum_{n_1=1}^{\infty}{n_1}^{1/2}\cos(n_1\theta)K_{1/2}(amn_1)+\sum_{n_2=1}^{\infty}K_0(\beta{mn_2})\nonumber \\
+\frac{2}{\beta}\sqrt{\frac{2\pi{a}}{m}}\sum_{n_1=1}^{\infty}\sum_{n_2=1}^{\infty}{n_1}^{1/2}\cos(n_1\theta)\frac{K_{1/2}\left(amn_1\sqrt{1+(2\pi{n_2}/{\beta}m)^2}\right)}{\sqrt[4]{1+(2\pi{n_2}/{\beta}m)^2}}\bigg]\label{mu-aT}, 
\end{eqnarray}

It is easy to verify the equivalence between (\ref{mu-bT}) and (\ref{mu-aT}) when $a\rightarrow\infty$ according to \cite{guida2}. 

To obtain the asymptotic expression for high temperatures, one can start summing over $n_1$ in (\ref{mu-aT}):

\begin{equation}
\sum_{n=1}^{\infty}n^{1/2}\cos(n\theta)K_{1/2}(nx)=\frac{1}{2}\sqrt{\frac{\pi}{2x}}\left(\frac{\cos(\theta)-e^{-x}}{\cosh(x)-\cos(\theta)}\right),
\nonumber
\end{equation}
while $\displaystyle\sum_{n_2=1}^{\infty}K_{0}(\beta{mn_2})$ can be rewritten by \cite{gradshteyn}: 
\begin{eqnarray}
\sum_{n=1}^{\infty}K_{0}(nx)=\frac{1}{2}\log\left(\frac{{\gamma}x}{4\pi}\right)+\frac{\pi}{2x}+{\pi}\sum_{p=1}^{\infty}\left[\frac{1}{\sqrt{x^2+(2{\pi}p)^2}}-\frac{1}{2\pi{p}}\right] , [x>0], \label{somaK0}
\end{eqnarray}
where $\gamma=1,781072...$ is the Euler constant, defined by 
${\log\gamma}=C=\displaystyle\lim_{s\rightarrow\infty}(\displaystyle\sum_{m=1}^{s}1/m-\log{s})=0,577215...$. 

Making $x=\beta{m}\ll{1}$ em (\ref{somaK0}), we have:  

\begin{eqnarray}
\frac{1}{\mu}{\approx}1+\frac{e^2}{12\pi^2}\bigg[\frac{\pi}{2\beta{m}}\left(\frac{\sinh(am)}{\cosh(am)-\cos(\theta)}\right)+\frac{1}{2}\log\bigg(\frac{\gamma\beta{m}}{4\pi}\bigg)
+\frac{1}{2}\sum_{n=1}^{\infty}\frac{1}{n}\left(\frac{\cos(\theta)-e^{-2\pi{a{n}/\beta}}}{\cosh(2\pi{an}/\beta)-\cos(\theta)}\right)\bigg], 
\end{eqnarray}

Calculating now the Casimir free energy of the system, given by $F=-\log{\cal Z}/\beta$, we have 

\begin{eqnarray}
F=-\frac{a{\ell}^2}{16{\pi}^2}\int_{s_0}^{\infty}\frac{ds}{s^3}(seB){\rm csch}(seB)e^{-sm^2}
\left[1+2\sum_{n_1=1}^{\infty}\cos(n_1\theta)e^{-(n_{1}a)^2/4s}\right]\left[1+2\sum_{n_2=1}^{\infty}e^{-(n_2\beta)^2/4s}\right]\label{energia-livre1}. 
\end{eqnarray}
Discarding terms which are proportional to $a\ell^2$ as spurious, since they represent a homogeneous vacuum energy density and considering 

\begin{equation}
x{\rm csch}(x)=\frac{2x}{e^x-e^{-x}}=\frac{2xe^{-x}}{1-e^{-2x}}=2x\sum_{n=1}^{\infty}e^{-(2n-1)x}; x\geq{0},\label{soma1} 
\end{equation}
the free energy becomes: 

\begin{eqnarray}
F=-\frac{a{\ell}^2eB}{4{\pi}^2}\sum_{n_1=1}^{\infty}\sum_{n_3=1}^{\infty}\cos(n_1\theta)\int_{0}^{\infty}\frac{ds}{s^2}e^{-(2n_{3}-1)eBs-m^2s-(n_1a)^2/4s}
\nonumber \\
-\frac{a{\ell}^2eB}{2{\pi}^2}\sum_{n_1=1}^{\infty}\sum_{n_2=1}^{\infty}\sum_{n_3=1}^{\infty}\cos(n_1\theta)\int_{0}^{\infty}\frac{ds}{s^2}e^{-(2n_{3}-1)eBs-m^2s-(n_1a)^2/4s-(n_2\beta)^2/4s}. 
\end{eqnarray}

The integrals of the above equation are representations of the Bessel function. 

\begin{eqnarray}
F=-\frac{a{\ell}^2eB}{{\pi}^2}\sum_{n_1=1}^{\infty}\sum_{n_3=1}^{\infty}\cos(n_1\theta)\bigg[\frac{\sqrt{m^2+(2n_3-1)eB}}{n_1{a}}K_{1}\left(n_{1}a\sqrt{m^2+(2n_3-1)eB}\right)
\nonumber \\
+2\sum_{n_2=1}^{\infty}\sqrt{\frac{m^2+(2n_3-1)eB}{(n_1{a})^2+(n_2{\beta})^2}}K_{1}\left(
\sqrt{[(n_{1}a)^2+(n_{2}\beta)^2][m^2+(2n_3-1)eB}]\right)\bigg]\label{energia-livre2}
\end{eqnarray}

For a strong magnetic field, $eB\gg{m^2}$, it is enough to consider $n=1$ in (\ref{soma1})since it is the term which gives the most relevant contribution. It corresponds to get only $n_3=1$ in (\ref{energia-livre2}). Doing this and writting  $m_B\equiv\sqrt{m^2+eB}$ \cite{guida3}, 

\begin{eqnarray}
F=-\frac{a{\ell}^2eB}{{\pi}^2}\sum_{n_1=1}^{\infty}\cos(n_1\theta)\bigg[
\frac{m_{B}}{n_1{a}}K_{1}\left(n_{1}am_{B}\right)+2\sum_{n_2=1}^{\infty}\frac{m_B}{\sqrt{{(n_1{a})^2+(n_2{\beta})^2}}}K_{1}\left(
m_{B}\sqrt{(n_{1}a)^2+(n_{2}\beta)^2}\right)\bigg]\label{energia-livre3}
\end{eqnarray}

Let's come back to (\ref{energia-livre1}) considering the case in which $B=0$, $\beta{=\infty}$ and $m=0$ from the equation, where $s\rightarrow{1/s}$

\begin{eqnarray}
F=-\frac{a{\ell}^2}{8{\pi}^2}\sum_{n=1}^{\infty}\cos(n\theta)\int_{0}^{\infty}{ds}{s}
e^{-s(na)^2/4}.  
\end{eqnarray}

Solving the above integral, 

\begin{eqnarray}
F=-\frac{2a{\ell}^2}{{\pi}^2{a^4}}
\sum_{n=1}^{\infty}\frac{\cos(n\theta)}{n^4}. 
\end{eqnarray}

The sum above can be rewritten using the $\varphi{_4}$ Bernoulli's polynomial 
\cite{charles-jordan}, 

\begin{equation}
\sum_{n=1}^{\infty}\frac{\cos(n\theta)}{n^4}={\pi}^4\left[\frac{1}{90}-\frac{1}{3}\left(\frac{\theta}{2\pi}\right)^2+\frac{2}{3}\left(\frac{\theta}{2\pi}\right)^3-\frac{1}{3}\left(\frac{\theta}{2\pi}\right)^4\right] ; 0<{\theta}<2\pi.\label{bernoulli} 
\end{equation}

Finally, the free energy is expressed as, 

\begin{eqnarray}
F=-\frac{2{\ell}^2{\pi}^2}{a^3}\left[\frac{1}{90}-\frac{1}{3}\left(\frac{\theta}{2\pi}\right)^2+\frac{2}{3}\left(\frac{\theta}{2\pi}\right)^3-\frac{1}{3}\left(\frac{\theta}{2\pi}\right)^4\right]; 0<{\theta}<2\pi. 
\end{eqnarray}

The Casimir pressure can be obtained from $P=-\partial{F}/\partial{V}$, where $V=a{\ell}^2$

\begin{eqnarray}
P=-\frac{6{\pi}^2}{a^4}\left[\frac{1}{90}-\frac{1}{3}\left(\frac{\theta}{2\pi}\right)^2+\frac{2}{3}\left(\frac{\theta}{2\pi}\right)^3-\frac{1}{3}\left(\frac{\theta}{2\pi}\right)
^4\right]; 0<{\theta}<2\pi. 
\end{eqnarray}

From the four roots of the polynomial, two are between $0$ and $2\pi$: 

\begin{equation}
\frac{\theta}{\pi}=\left\{\left[1-\sqrt{1-\frac{4}{\sqrt{30}}}\right]; 
\left[1+\sqrt{1-\frac{4}{\sqrt{30}}}\right]\right\}{\approx}
\{0,48067; 1,51933\}\nonumber
\end{equation}

Between $0$ and the first root, the polynomial is positive and between the first and the second it is negative. So the Casimir's pressure changes signal but in the opposite way to the $\varphi_4$ Bernoulli polynomial.

Summarizing our results, eq.(8) gives the change in the magnetic permeability due to the 
combination of confinement, temperature effects and the $\theta$ deformation parameter. When $\theta=0$, we have a diamagnetic permeability arising from a periodic boundary condition. When $\theta=2\pi$, the anti-periodic boundary condition is present and the permeability turns to a paramagnetic one. Eq. (18) gives the exact expression for the free energy of the system and all the limits that can be taken will give results compatible with the current literature \cite{guida2, guida3}.

}   

\end{document}